\documentclass[aip,rsi,reprint,graphicx,twocolumn]{revtex4}
\usepackage{epsfig}
\usepackage{graphicx}% Include figure files
\usepackage{dcolumn}% Align table columns on decimal point
\usepackage{bm}% bold math
\usepackage{lmodern}
\usepackage{textcomp}
\usepackage{caption}
\usepackage{graphicx}
\usepackage{array}
\usepackage{amsmath}
\usepackage{array}
\usepackage{booktabs}
\usepackage{rotating}

\usepackage{adjustbox}
\usepackage{blindtext}
\usepackage{graphicx}
\usepackage{array}
\usepackage{amsmath}
\usepackage{array}
\usepackage{booktabs}
\usepackage{rotating}

\usepackage{setspace}

\usepackage{adjustbox}
\usepackage{blindtext}
\usepackage{epsfig}
\usepackage{graphicx}% Include figure files
\usepackage{dcolumn}% Align table columns on decimal point
\usepackage{bm}% bold math

\usepackage{xcolor}

\usepackage[paperwidth=250mm,paperheight=330mm,centering,hmargin=1.5cm,vmargin=1.5cm]{geometry}

\begin{document}
%\preprint{APS/123-QED}

\title{ Realization of Planar Optical Tweezer (2D-LOT) Powered by Light Sheet }% Force line breaks with \\   % and Drug Evaluation / Treatment Studies

%\title{ Planar Optical Tweezer Trap (2D-LOT) System Realized by Light Sheet Illumination $\&$ Orthogonal Widefield Detection }% Force line breaks with \\   % and Drug Evaluation / Treatment Studies

\author{ Neptune Baro$^1$ and Partha Pratim Mondal$^{1,2}$  }
 %\altaffiliation{}%Lines break automatically or can be forced with \\
 %\email{partha@fisica.unige.it}
%\email[Corresponding author: partha@iisc.ac.in;  \\ $\dag$First two authors contribute equally to this work.]{}
\affiliation{%
$^1$ Department of Instrumentation and Applied Physics, Indian Institute of Science, Bangalore 560012, INDIA
}%
\affiliation{%
$^2$ Centre for Cryogenic Technology, Indian Institute of Science, Bangalore 560012, INDIA
}%
\date{\today}% It is always \today, today,
%\doublespacing   

%\usepackage{lipsum}

\makeatletter
\renewcommand\frontmatter@abstractwidth{\dimexpr\textwidth-1in\relax}
\makeatother

\begin{abstract}

We report the realization of the first planar optical tweezer trap system by a sheet of light. To visualize the trapping of the target object (dielectric bead or live cell) in a plane, an orthogonal widefield detection is employed. The planar / two-dimensional lightsheet optical tweezer (2D-LOT) sub-system is realized in an inverted microscopy mode with illumination from the bottom. A 1064 nm laser (power $\sim 500 mW$) is expanded and directed to a combination of cylindrical lens and high NA objective lens to generate a tightly-focused diffraction-limited light sheet. The object to be trapped is injected in the specimen chamber (consists of two coverslips placed at a distance of $\approx 1~mm$) using a syringe. The solution containing the objects stayed in the chamber due to the surface tension of the fluid. The illumination of trap-laser light is along Z-direction (with coverslip along XZ-plane) whereas, the detection is achieved perpendicular to the coverslip (along Y-axis). The orthogonal detection  is employed to directly visualize the trapping in a plane. To better visualize the specimen, a separate white light illumination sub-system is used. The characterization of system PSF estimates the size of light sheet trap PSF to be, $2073.84 ~\mu m^2$ which defines the active trap region / area. Beads are tracked on their way to the trap region for determining the trap stiffness along Z and X i.e, $k_z = 1.13 \pm 0.034 ~pN/\mu m$ and $k_x = 0.74 \pm 0.021 ~pN/ \mu m$. Results (image and video) show real-time trapping of dielectric beads in the trap zone (2D plane) generated by the light sheet. The beads can be seen getting trapped from all directions in the XZ-plane. Prolonged exposure to the light sheet builds up a 2D array of beads in the trap zone. Similar experiments on live NIH3T3 cells show cells trapped in the 2D trap. The potential of the planar trap lies in its ability to confine objects in two dimensions, thereby opening new kinds of experiments in biophysics, atomic physics, and optical physics.  \\

{\bf{Statement of Significance:}} The ability to trap and confine objects in two dimensions / a plane is an incredible feat that paves the way for new experiments in physical and biological sciences. Currently, no technique can achieve two-dimensional trapping of objects. Such a trap is realized by a sheet of light generating a near-rectangular potential trap-zone, enabling tweezing confined to a plane. This is unlike existing point-focus-based tweezers that are capable of point-potential well. The technique is expected to have widespread applications in science and engineering.  \\

Corresponding author (Partha Pratim Mondal): partha@iisc.ac.in  \\

%{\bf{ORCID ID:}} 0000-0003-0241-0081   \\

\end{abstract}

\maketitle

\clearpage

\section{Introduction}

The manipulation of objects in a 2D / plane is fascinating. An optical tweezer system that is capable of trapping objects in a plane has applications across the wide spectrum of science and engineering. The existing tweezer systems that are primarily point-traps can trap and study a single object at any given point in time. Hence, studies requiring many particle / object interactions are not possible. However, 2D tweezer traps can elevate the constraint faced by point-tweezers and may enable trapping several particles (e.g., atoms / molecules / cells, etc.) simultaneously in a single plane, thereby giving an opportunity to study their interactions and collective dynamics constrained to two dimensions (2D physics). Moreover, a tweezer trap system that can enable studies limited to two dimensions is of utmost need for advancing touch-free manipulation. \\

The first ever optical tweezer was realized by Arthur Askhin in the year 1970 \cite{ash1970}. The technique was successfully used to trap both living and non-living objects with sizes as small as a few microns (dielectric beads, bacteria, red blood cell) to as large as a few millimeters (multicellular organisms) \cite{ash1986} \cite{ash1987}. Subsequently, the technique has grown into a major investigation tool giving it a multidisciplinary nature. For example, the technique is used to investigate, short-range colloidal interactions, and cellular-liposome interactions \cite{spy2014}. Moreover, the technique is used for manipulating microscopic objects such as rotation of microscopic objects, liposome biomechanics, and single molecule force spectroscopy \cite{rog2014} \cite{bian2006}  \cite{bust2020} \cite{zalt}. To carry out diverse studies involving forces in the pico-Newton range, several important variants of point-trap have emerged over the years \cite{stig} \cite{cecc} \cite{mail} \cite{tam} \cite{avs}. This includes ring-vortex traps, shape-phase holography, scanning point traps, beam shaping, and hyperbolic meta-materials \cite{lee2010} \cite{shan2011} \cite{roi2006} \cite{fau1995} \cite{woe2013} \cite{rod2013} \cite{paral2024}. In the year 2022, our group proposed the first light sheet based optical tweezer (LOT) that demonstrated capturing objects in a line \cite{LOT}. This was the first time a non-traditional point-based PSF was used to trap objects. However, unlike the existing point-based traps, the light sheet PSF has the ability to trap objects in a plane. Such a capability has the ability to push the boundaries of science and engineering and may enable new kinds of studies constrained to two dimensions or a plane including touch-free immobilization of freely moving objects (living and non-living). \\  

To circumvent the limitations of existing point-trap-based optical tweezers and enable new capabilities, we demonstrate a new kind of optical trap powered by lightsheet. The objects (dielectric beads / cells) are trapped in a sheet of light representing the trapping plane, and a single layer of object is realized. Since trapping is limited to two dimensions, this allows the study of particle-particle interaction in a single layer. In principle, the technique may allow layer-by-layer deposition of objects (atoms, single molecules, dielectric beads, cells, etc.) in a controlled manner to construct a volume or the study of a single layer of particles (e.g., Graphene layers). In biology, the technique can be used to arrange cells in a single layer and many such layers can be put together to artificially construct a tissue. Many such applications are possible in diverse research areas. Overall, 2D-LOT system is expected to advance biological, physical, and engineering sciences.       

\begin{figure*}
	\includegraphics[width=15cm]{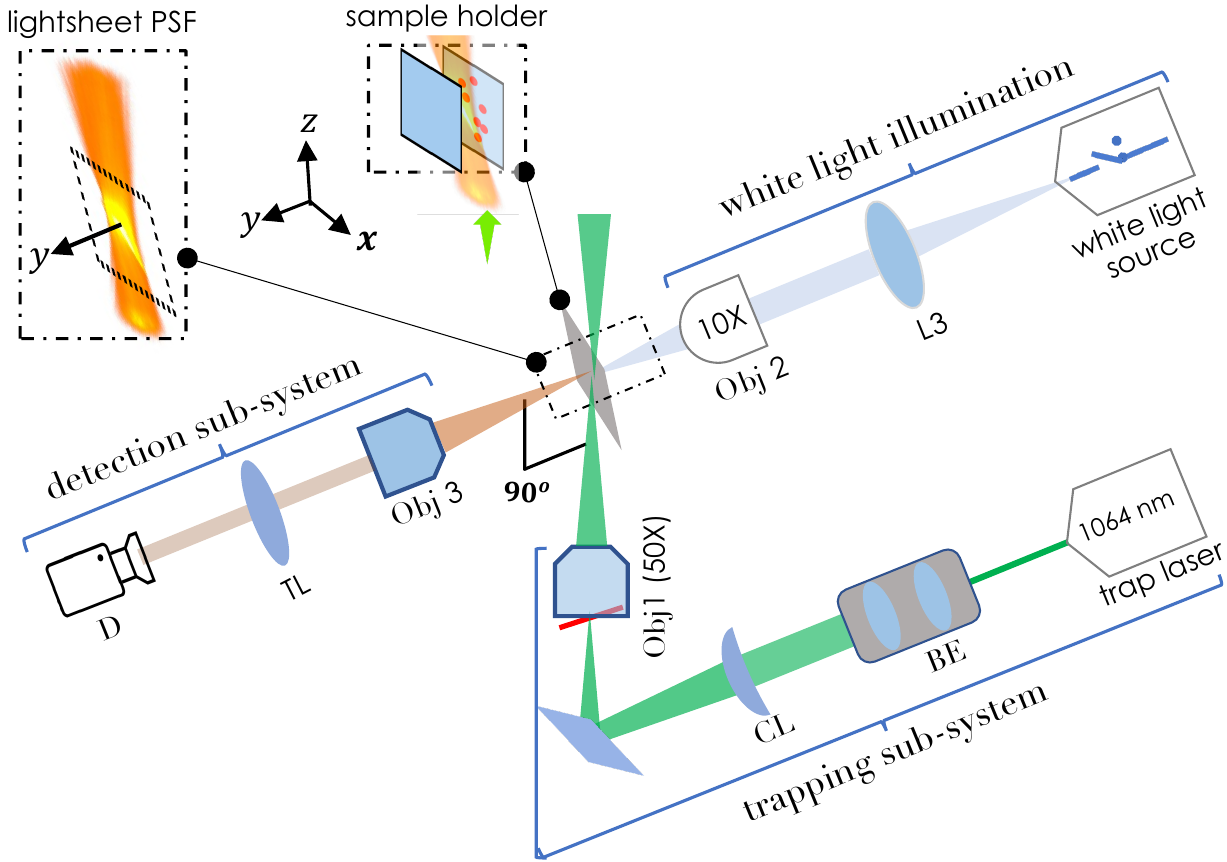}
	%	\vspace{-0.3cm}
	\caption{{\bf{ Schematic diagram of planar 2D-LOT system :}} The system consists of a lightsheet trapping sub-system, a sample-holder, a white light illumination system, and a detection system. The trapping sub-system uses a $1064 ~nm$ IR laser along with a combination of cylindrical lens (CL) and objective lens (Obj1, 50X) lens for generating a diffraction-limited lightsheet PSF to realize 2D planar optical trap. A separate white light illumination system (consists of, white LED light, 10X objective lens (Obj2), and a biconvex lens (L3)) is used to visualize the specimen. The detection (D) is achieved by an orthogonal detection system (placed perpendicular to the trapping sub-system) using a third objective (Obj3) and tube lens (TL) to directly visualize objects being trapped in the lightsheet PSF. The insets show lightsheet PSF trapping beads in the sample-holder and the detection along $y$-axis. } 
	\label{fig:par1}
\end{figure*}

\begin{figure*}
	\includegraphics[width=22cm]{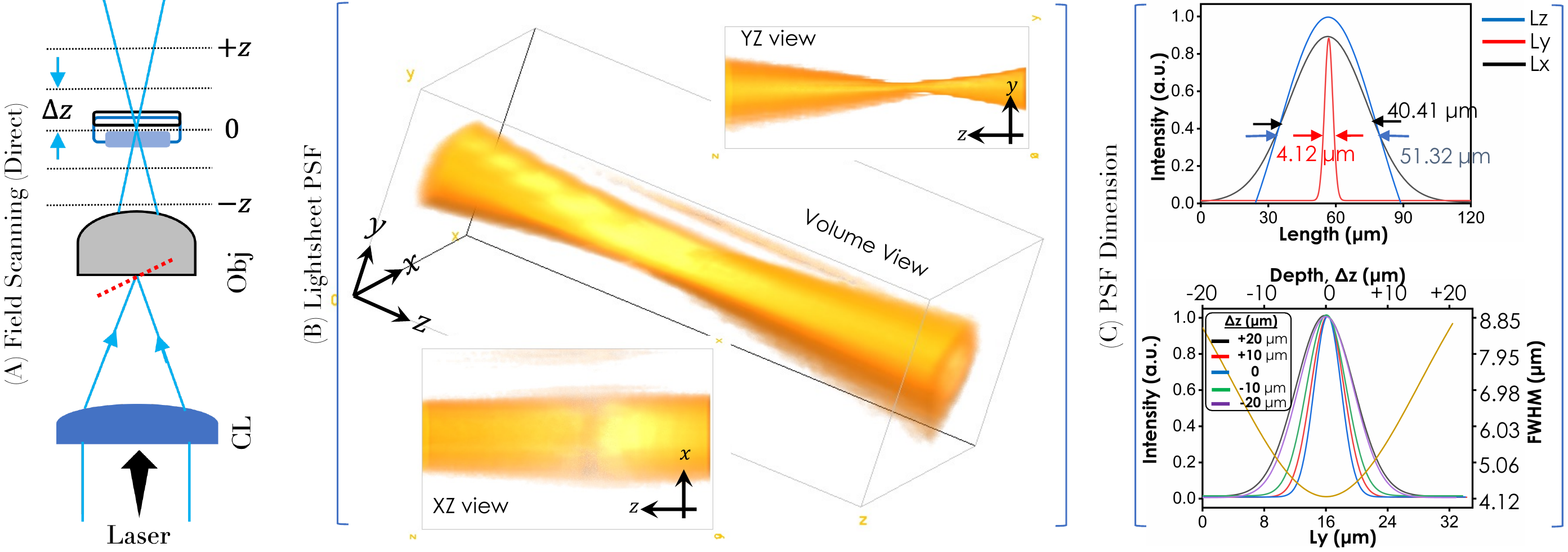}
	%	\vspace{-0.3cm}
	\caption{{\bf{ Light Sheet PSF Characterization:}} (A) Schematic of the optical setup used for recording the light field at and about the focal plane. (B) The 3D lightsheet field was obtained by stacking all the recorded intensity images at regular $Z$-interval of $2 ~\mu m$. Alongside two different views (XZ and XY) of the field are also shown. (C) Light intensity plots determine the size of the sheet of light to be, $51.32 \mu m \times 40.41 \mu m $ ($L_z \times L_x$) with a thickness ($L_y$) of $4.12 ~\mu m$. In addition, the thickness (measured in terms of FWHM) of the lightsheet PSF at different distances from the focus ($-20\mu m, ~-10\mu m, ~0, ~+10\mu m, ~+20\mu m$) is also calculated. } 
	\label{fig:par1}
\end{figure*}

\begin{figure*}
	\includegraphics[width=21cm]{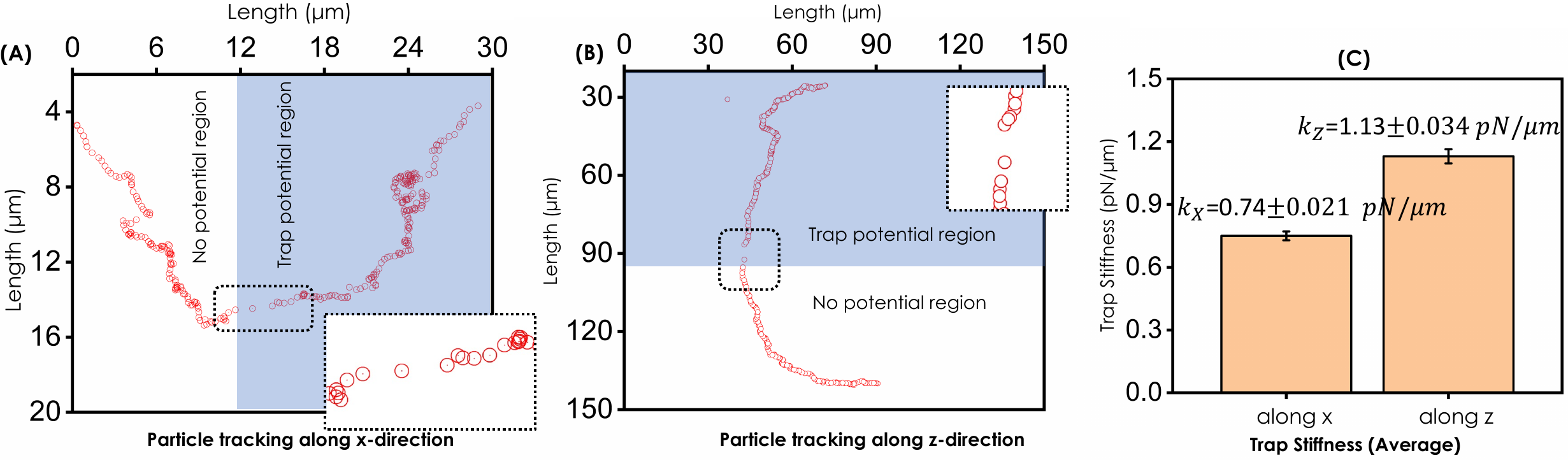}
	%	\vspace{-0.3cm}
	\caption{{\bf{ Trap Stiffness along X and Z Direction in the Trap-zone:}} (A) A typical dielectric bead tracked as it crosses from free (no-potential) region to trap-potential region /zone along, (A) X-direction, and (B) Z-direction. (C) Average value of trap stiffness ($k_x = 0.74\pm 0.021 ~pN/\mu m$, and $k_z = 1.13\pm 0.034 ~pN/\mu m$) obtained from tracking $5$ different particles as it cross-over from free-zone to trap-zone. } 
	\label{fig:par1}
\end{figure*}

\section{Results}

\subsection{2D-LOT Trap System}

The schematic diagram of the developed $2D-LOT$ system is shown in Fig. 1. The system consists of a trapping sub-system, a sample holder, a bright light illuminator, and a $4f$ widefield detection system. The trapping sub-system uses a $1064~nm$ laser light source, a $4X$ beam-expander, a cylindrical lens ($f=75 ~mm$), and an objective lens (0.5 NA, 50X). These components along with other optics (mirrors, iris, etc.) generates a tightly focused sheet of light. The sample holder is a pair of coverslips separated by $2 ~mm$ where the sample is injected using a syringe. The illumination consists of a broadband white light LED, and a low NA objective lens (10X, 0.25 NA) is used to illuminate the specimen. The detection sub-system is orthogonal to the trapping sub-system and faces the lightsheet (along Y-axis) which is perpendicular to the lightsheet plane (XZ) (see, the inset in Fig. 1). Detection is essentially a 4f optical configuration that consists of an objective lens (20X, 0.4 NA), tube lens ($f=125~mm$) and a fast CMOS camera (Gazelle, Pointgray, USA)). The dielectric beads trapped in the light sheet plane (XZ) are directly visualized in the camera.  

\subsection{Calibration of 2D-LOT System and Lightsheet PSF Characterization }

The light sheet PSF generated by the trapping sub-system needs to be characterized for quality trapping. Fig. 2A shows the schematic diagram of the optical setup used for characterizing the lightsheet PSF. A CCD camera is placed in the beam-path and scanned about the focus of the objective lens. The data (image of the field) is recorded at a sampling distance of $2 ~\mu m$ along the z-axis. The images are then stacked together to reconstruct the 3D field. The corresponding volume view along with the sectional views (XZ and YZ) of the light sheet field is shown in Fig. 2B. Thus the resultant sheet of light generated by the trapping sub-system is quite apparent. In addition, the thickness of the field (measured in terms of FWHM along Y-azis) is determined at varying distances along beam propagation ($z$-direction) as shown in Fig. 2C (intensity plots). This characterizes the spread of lightsheet field along the propagation direction. Finally, intensity plots are carried out to determine the dimension of the light sheet PSF or the effective trapping region / zone. The size of trap-zone realized by the light sheet is calculated to be, $40.41~\mu m \times 51.32 ~\mu m$ ($L_x \times L_z$) in XZ-plane. This is the effective planar zone that can be used for trapping the objects (dielectric bead / cell).            

\subsection{Trap Stiffness along X and Z Axes}

Efficient trapping depends on several factors including the strength of the optical field. Since 2D-LOT is a two-dimensional trap, the stiffness needs to be calculated along both $X$ and $Z$ axis. To calculate the stiffness values, we have used a technique similar to that in LOT \cite{LOT}, where camera frames / video are used to identify sudden pulls caused by the 2D trapping potential zone. In general, the beads exhibit free Brownian motion in the medium. Upon turning ON the trap laser at $t=0$, causes the beads to move towards the trap-zone. The journey from time $t=0$ to the 2D trap and eventually to the trap potential zone is recorded by a high-speed CMOS camera (Gazelle, Pointgray, USA). The time and the distance traveled by the bead for the entire journey are calculated from the number of frames recorded. From the video, several free beads are identified and tracked to the trap-zone.  \\

Near the trap-zone, the force can be approximately modeled by Hooke’s law, and the restoring/gradient force is given by, $F_G = -kx$, where $k$ is the trap stiffness $(N/m)$ and $x$ is the distance. The other force acting on the bead during its directed motion to the potential zone is the viscous force experienced by the bead moving through the medium, i.e, $F_V = 6\pi \eta r_b v$ where, $\eta$, $r_b$ and $v$ are the medium viscosity, bead radius (assuming spherical) and velocity, respectively. The bead experiences these forces (gradient force and viscous drag force) which are opposite to each other. Balancing these forces produces, $k=6\pi \eta r_b / t$, where, $t=x/v$  with $x$ as the displacement of the bead and $t$ as the time taken for the journey. Note that, this is an approximate formula where the effect of gravity is neglected thereby ruling out forces due to weight and buoyancy. \\

From the particle-track plots, the beads approaching the boundary (between NO-potential and Trap-potential zone) undergo sudden pull towards the trap-zone. Once inside the zone, the beads do not feel any force due to negligible intensity-difference, except the Brownian motion exhibited by the beads. The time-taken during the process i.e., the particle crossing the boundary and ultimately reaching to the potential zone determines the trap stiffness. This is true for particles crossing from both X and Z directions for which the trap stiffness is $k_x$ and $k_z$. During the process, the dielectric beads are tracked as they move from the free-zone to the trap-zone (see, {\bf{Supplementary Video 1}}).  Fig. 3A and 3B show a typical dielectric particle being suddenly trapped by the trap zone while crossing the boundary along X and Z directions, respectively. The time-taken for the journey is determined from the number of frames ($4$ frames along $X$ and $2$ frames along $Z$) which is calculated to be, $200 ~s$ and $100 ~s$ for $X$ and $Z$, respectively. For our case, the beads are assumed spherical and have a diameter of $2~\mu m$, and a density of $\sim 2000 ~kg/m^3$. So, the mass of the bead is, $m=\rho V$ where the volume of the bead can be calculated using the formula, $V=4/3 \pi (d/2)^3$. Knowing the viscosity of deionized water to be $\eta=0.8925 \times 10^{-3} ~Pa~s$, the approximate trap-stiffness is calculated to be, $k_x = 0.74 ~pN/\mu m$ along X-axis and $k_z = 1.13 ~pN/ \mu m$ along Z-axis.    

\begin{figure*}
	\includegraphics[width=22cm]{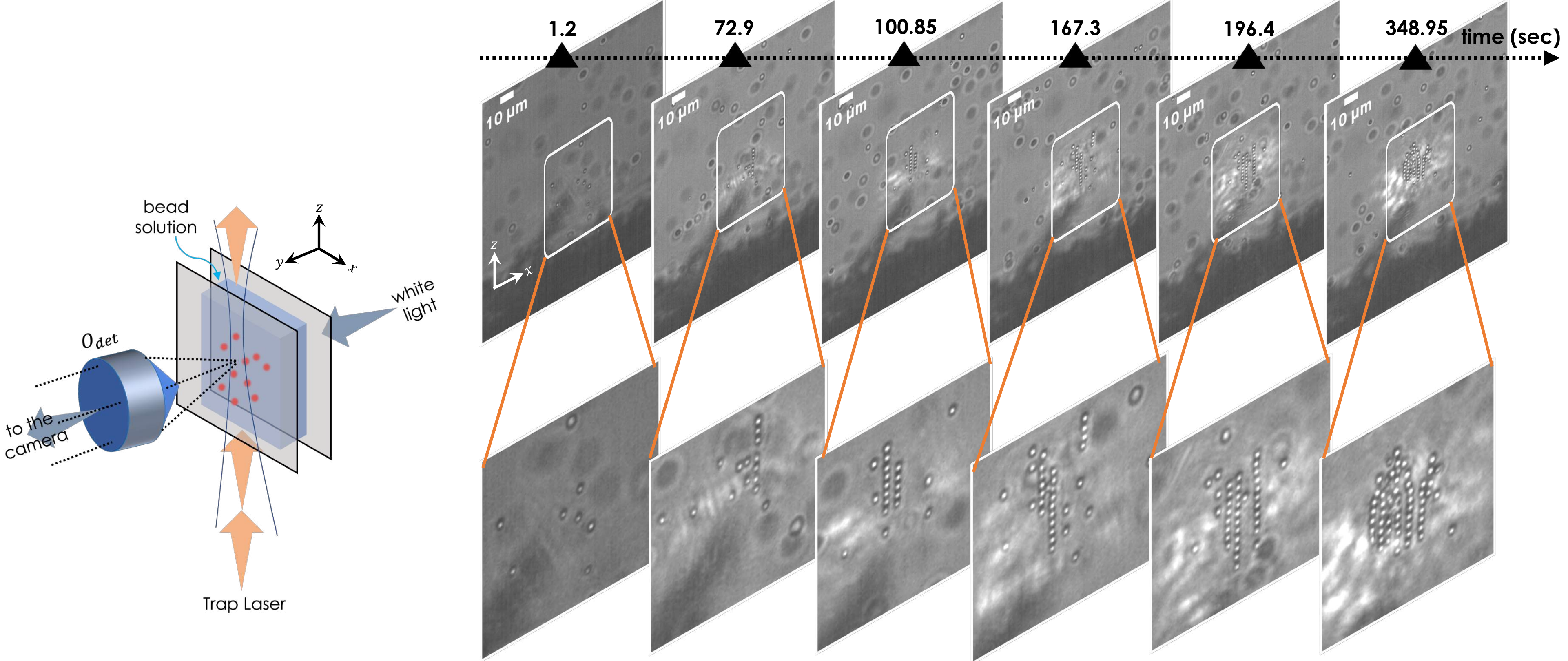}
	%	\vspace{-0.3cm}
	\caption{{\bf{ Trapping Dielectric Beads in a Lightsheet Plane:}} A drop of dielectric beads suspended in deionized water is injected in a closely spaced sample holder (see, schematic diagram). The sample is then subjected to a sheet of light causing a trap potential region / zone (marked by a white rectangle). The beads (size $\sim 2 \mu m$) are attracted by the potential. Over time (from $1.2-348.95 ~secs$), a single 2D layer of beads appears to form. The enlarged region is also shown. Scale bar $=10 ~\mu m$ } 
	\label{fig:par1}
\end{figure*}

\begin{figure*}
	\includegraphics[width=22cm]{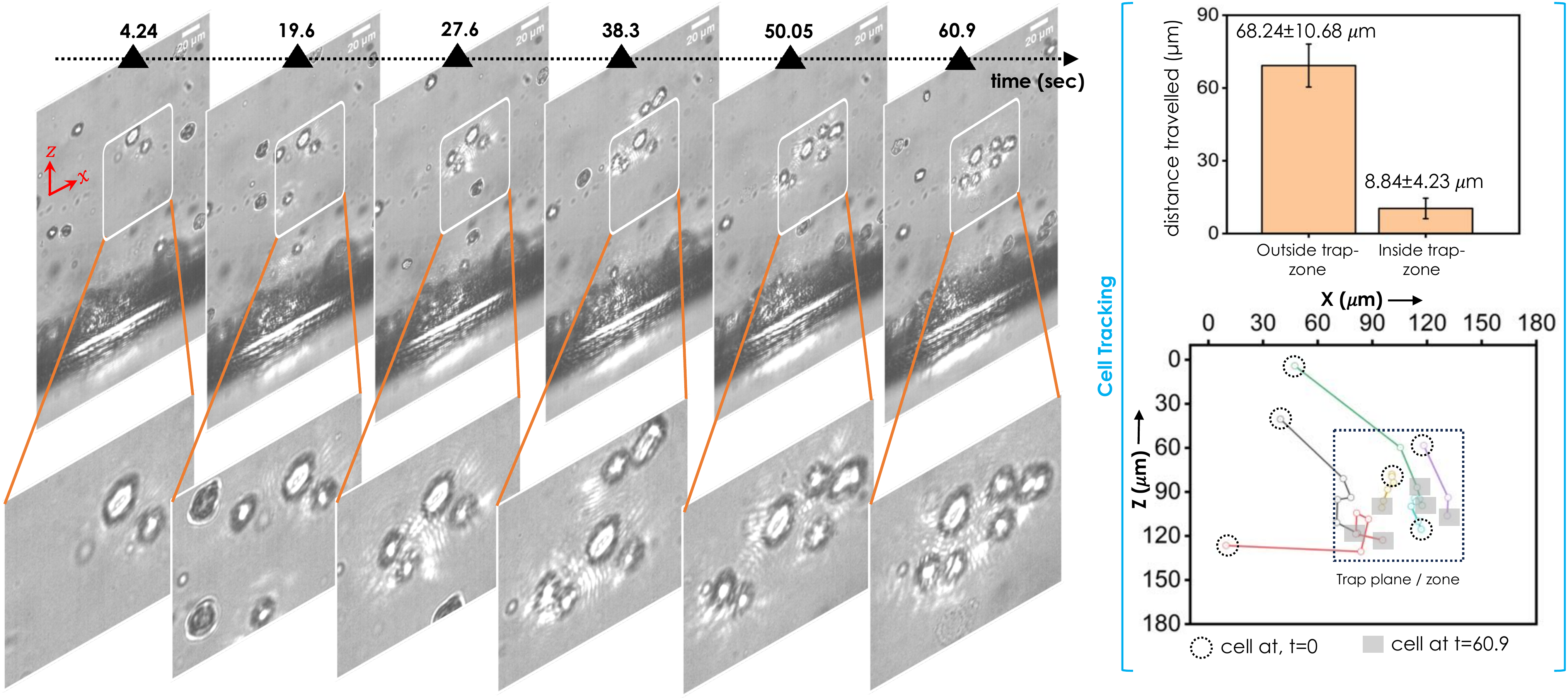}
	%	\vspace{-0.3cm}
	\caption{{\bf{ Trapping NIH3T3 cells in a Lightsheet Plane:}} NIH3T3 cells suspended in the medium are subjected to the sheet of light (marked by a white rectangle). With time, the cells get randomly trapped in a 2D plane (realized by the sheet of light). The enlarged region along with time is also shown. Alongside, track plot and distance bar-plot are also shown for $t=4.24 - 60.9 ~secs$. The long-distance ($68.24 \pm 10.68~\mu m$) indicates sudden pull by the trap-zone, whereas the short distance ($8.84 \pm 4.23 ~\mu m$) within the trap-zone suggests Brownian motion exhibited by the object (cell). Scale bar $=20 ~\mu m$ } 
	\label{fig:par1}
\end{figure*}

\subsection{Trapping Dielectric Beads and NIH3T3 Cells in a Plane}

To validate the working of 2D-LOT system, two different objects are trapped in a plane. We choose to trap microscopic dielectric silica beads (spherical, size $\sim 2 \mu m$) suspended in deionized water, and live NIH3T3 cells (nearly spherical, size $10-17 ~\mu m$) in a cell medium. Fig. 4 shows the trapping of beads in an effective potential zone / region generated by lightsheet PSF (see, white rectangle). The size of the trap-zone is approximately, $L_z \times L_x = 51.32~\mu m \times 40.41 ~\mu m = 2073.84 \mu m^2$ in ZX-plane. To trap beads, the trap-laser is turned ON, and the video of the entire trapping process is recorded (see, {\bf{Supplementary Video 1}}). We have chosen to show images at a few chosen time-points beginning from $t=1.2 - 348.95 ~secs$. At time $t=1.2~secs$, only $4$ beads appear to be in the trapping zone. The subsequent frame at $t=72.9$ shows more beads trapped in the zone. With increasing time, several beads ($>50$) are seen trapped in the trap-zone. Finally, at $t=348.95$, the beads can be seen trapped in a single layer / plane. Turning off the trap laser shows a sudden fall of dielectric beads. The entire process is captured in {\bf{Supplementary Video 1}}.  This conclusively shows the two-dimensional trap realized by light sheet PSF. \\

Subsequently, we have chosen to work with live specimens such as, live NIH3T3 cells as shown in Fig. 5. In the initial frame at $t=4.24 ~secs$, only two cells are seen trapped in the light sheet potential zone. Subsequent frames show an increase in the number of cells trapped in the potential zone, with a maximum of $6$ cells at $t=60.9 ~secs$. Compared to dielectric beads, the number of trapped cells is less which can be attributed to its large size (approximately $7$ times larger than the beads). In addition, the cells were tracked on their way to the trap-zone. It is observed that the distance traveled by the cell from the initial position (at $t=4.24$ just outside trap-zone) to the trap-zone is longer ($\sim 68.24 \pm 10.68~\mu m$) as compared to displacements ($\sim 8.84 \pm 4.23 ~\mu m$) within the trap-zone (see, time-points, $t>4.24$ in Fig. 5). Once the cells are trapped in the trap-zone, they continue to exhibit Brownian motion as noted from the short displacements between time-points. This manifests the sudden force felt by the cell close to trap-zone. At the end ($t=60.9 ~secs$), all the cells were clearly seen trapped in a single plane / trap-zone, realized by lightsheet PSF. The entire process of cell getting trapped in the trap-zone is shown in {\bf{Supplementary Video 2}}. Trapping of live cells in a chosen 2D plane opens up new possibilities, both in biophysics and cell biology.

\section{Conclusion $\&$ Discussion}

In this article, we report the first two-dimensional / planar optical trap realized by a sheet of light. The technique uses a light of $1064~nm$ light along with a combination of cylindrical lens and objective lens to generate a tightly-focused diffraction-limited light sheet. The system is realized in an orthogonal optical configuration with detection sub-system perpendicular to the trapping sub-system. This helps in direct visualization of trapping in the light sheet plane. \\  

Trapping objects in two dimensions / plane is an incredible development that brings in new possibilities both in physical and biological sciences. To achieve this feat, a new kind of optical tweezer-trap system (called 2D Lightsheet Optical Tweezer or 2D-LOT) is developed. The technique allows layer-by-layer deposition, plane-by-plane interrogation, and many particle / object interaction in a plane. Unlike existing point-based tweezers, the developed system allows trapping objects in a 2D plane. The technique is used to optically trap objects such as dielectric beads and even live specimens (NIH3T3 cells) in a plane, allowing the understanding of their physical / biological interactions in a single layer / plane. \\        

The PSF is capable of trapping several objects in a plane (sheet of light). The light sheet can be better visualized in the volume view (see, Fig. 2). The size of the light sheet closely resembles a rectangle with a dimension of $51.32 \times 40.41 ~\mu m^2$ in the ZX-plane. Unlike the existing point-traps, the stiffness has two components for the planar trap, with $k_z$ along the propagation direction and $k_x$ along $X$ in the ZX/lightsheet-plane. To determine the trap-stiffness, tracking video based technique is employed where, particles are tracked from their initial free position to the particle crossing over to the potential zone (see, Fig. 3). Near the trap zone, the particle gets suddenly pulled due to difference in intensity, causing a directed movement to the trap zone. From the recorded frames, the distance covered and the time taken are determined, giving an approximate estimate of trap stiffness, $k_z = 1.13 \pm 0.034 ~pN/\mu m$ and $k_x = 0.74 \pm 0.021 ~pN/ \mu m$ along $Z$ and $X$ axes. This clearly shows slightly different stiffness along the propagation (Z) axis and other (X) axis. \\     

Two different objects are considered to evaluate the working of 2D-LOT system: dielectric beads and live NIH3T3 cells. The functioning of the system is demonstrated in brightfield mode (where the trapping is achieved by trap-laser and detection is carried out perpendicular to the trapping sub-system using a separate white light illumination (see, Fig. 1). The beads are trapped at different time-points beginning from its free position (outside trap-zone) to its travel to the trap-zone (see, Fig. 4). The same experiment is carried out for live NIH3T3 cells (see, Fig. 5). It is noted that the cells take relatively large time to enter the trap-zone as compared to dielectric beads, which is predominantly due to its large size. The corresponding recorded video for the dielectric bead and live NIH3T3 cells are shown in {\bf{Supplementary video 1 and 2}} that captures the entire tapping process. \\   

We anticipate that the new capability that allows the trapping of objects in a 2D/plane (generated by the dimension of the light sheet) has potential applications in science and engineering. In principle, the technique allows touch-free manipulation of objects or the study of their dynamics in two dimensions, discrete interrogation of an entire layer of single particles, and understanding the interactions between the objects (dielectric beads / cells / atoms / molecules) in a 2D/plane.

\section{Materials $\&$ Methods}

\subsection{Sample Preparation}

{\bf{Silica Beads:}} Silica dielectric beads in a $5 \%$ concentration of 5ml aqueous solution were purchased from Sigma Aldrich, Germany. The bead size is $2$ microns with a standard deviation of $0.2$ microns. The beads are diluted in deionized water to $1/10~th$ of the original concentration for carrying the trapping experiment. \\

{\bf{NIH3T3 Cells:}} NIH3T3 cells (mouse fibroblast cell line) were cultured in Dulbecco’s modified Eagle’s medium (DMEM) (Gibco™, Thermo Fisher Scientific, Waltham, MA, USA) supplemented with $10\%$ fetal bovine serum (FBS) and $1\%$ of penicillin-streptomycin. The cells were seeded at a density of $10^5$ cells in 35 mm dishes and maintained at 37 $^\circ$C in a humidified incubator with a $5\%$ CO2 atmosphere for 24 hrs. The dishes were observed for $70-80 \%$ confluency and washed twice with 1X PBS to remove debris. Subsequently, the cells were detached from the dishes using a diluted trypsin solution ($1\%$ trypsin in PBS) and collected in a 1.5 ml Eppendorf tube, followed by centrifugation for 2 minutes at 3000 RPM. The pelletized cells were resuspended in 1 ml of cell medium and used for trapping experiments. 

\subsection{ Image Acquisition and Data processing }

A CMOS (Gazelle, Pointgray, USA) camera is employed with a 125 mm tube lens to acquire the brightfield images. The graphical programming environment, Labview (NI instruments), is interfaced with the CMOS camera to control the data acquisition. The camera is adjusted to a suitable exposure time for crispy images and is set to a $550~px \times 1088~px$ window size for faster acquisition. Subsequently, $360~secs$ and $65~secs$ videos were captured at a frame rate of 70 fps to visualize the trapping of dielectric beads and Cells, respectively. 

\subsection{Tracking of Dielectric Beads and NIH3T3 cells tracking }

In general, tracking an object is performed in two steps -  1) Detection and identification of the object's position, 2) linking of detected objects from frame-to-frame to estimate their trajectories. Cells and Beads underdoing Brownian motion were tracked using Trackmate GUI available in Fiji. The DoG (Difference of Gaussian ) and Simple LAP tracker modules within TrackMate are used for the detection of the cells in individual frames and linking them, respectively. To track the cells, frames were extracted from the trapping video with an average time of $12.77$ secs between consecutive frames. The acquired frames then undergo background subtraction for better detection. The location of cells on each frame is generated by setting an estimated object diameter of $10 ~\mu m$ and using a threshold of $0.5$ in DoG filter. Post-detection, the cells need to be linked from frame to frame to acquire the entire track.  For this, the LAP tracker (with a linking max distance $=80~\mu m$,  gap closing max distance $=10 ~\mu m$, gap closing max frame $=2~frames$) was used. \\

{\bf{Detection and Identification using DoG Filter Module:}} The detection is based on the DoG filter. In this method, an approximate expected particle diameter (d) is fed to the module.  It generates two Gaussian filters with standard deviations, $\sigma_1, ~\sigma_2$ (with $\sigma_2 > \sigma_1$) which are used to filter the image stack. This is followed by subtraction, giving a smoothened image with sharp local maximums at particle locations. Each of these maximum acts as a detection spot. This spot is assigned a quality feature by taking the smoothened image value at the maximum. Between two spots with an expected object radius less than $d/2$, the one with the lowest quality is discarded. \\

{\bf{Linking of Detected Objects Using LAP Tracker:}} The linking of detected objects is determined in terms of linking cost. The linking of a spot $i$ (representing a detected object) at frame $t$ and spot $j$ in the consequtive frame $t+1$ has a cost. This cost is proportional to the squared distance between the linked spots. In the LAP tracker, the costs of all links between the two frames are minimized as a total and then the objects are linked between the frames. Additionally, links are discarded if the mean object intensity differs largely between two frames. An allowed size of the gap, (say, $2$ frames) specifies if the objects in the subsequent frames appear or disappear. \\

{\bf{Acknowledgements:}} The authors acknowledge support from parent institute (Indian Institute of Science, Bangalore, India). \\

{\bf{Contributions: }} PPM conceived the idea. NB, PPM carried out the experiments. NB prepared the samples. NB and PPM analyzed the results and prepared the figures. PPM wrote the paper by taking inputs from NB. \\

{\bf{Data Availability:}} The data that support the findings of this study are available from the corresponding author upon request. \\

{\bf{Supplementary Info:}} The manuscript is supported by two supplementary videos.  \\

{\bf{Disclosures:} }The authors declare no conflicts of interest. \\

\end{document}